\documentclass[preprint,showpacs,preprintnumbers,amsmath,amssymb]{revtex4}

\usepackage{graphicx}% Include figure files
\usepackage{dcolumn}% Align table columns on decimal point
\usepackage{bm}% bold math

\begin{document}

%\preprint{APS/123-QED}

\title{High-Capacity Quantum Associative Memories}% Force line breaks with \\

\author{M. Cristina Diamantini}
%\altaffiliation[Also at ]{Theory Division, CERN, CH-1211 Geneva 23, Switzerland}%Lines break automatically or can be forced with \\
\email{cristina.diamantini@pg.infn.it}
\affiliation{%
INFN and Dipartimento di Fisica, University of Perugia, via A. Pascoli, I-06100 Perugia, Italy
}%

%\author{Pasquale Sodano}
%\altaffiliation[Also at ]{Perimeter Institute of Theoretical Physics 31, Caroline St. North, Waterloo, Ontario N2L2Y5, Canada}%Lines break automatically or can be forced with \\
%\author{Second Author}%
%\email{pasquale.sodano@pg.infn.it}
%\affiliation{%
%INFN and Dipartimento di Fisica, University of Perugia, via A. Pascoli, I-06100 Perugia, Italy
%}%

\author{Carlo A. Trugenberger}
%\altaffiliation[Also at ]{Theory Division, CERN, CH-1211 Geneva 23, Switzerland}%Lines break automatically or can be forced with \\
%\author{Second Author}%
\email{ca.trugenberger@bluewin.ch}
\affiliation{%
SwissScientific, chemin Diodati 10, CH-1223 Cologny, Switzerland
}%

%\author{M. Cristina Diamantini}
%\homepage{http://www.Second.institution.edu/~Charlie.Author}
%\affiliation{
%Second institution and/or address\\
%This line break forced% with \\
%}%

\date{\today}% It is always \today, today,
             %  but any date may be explicitly specified

\begin{abstract}
We review our models of quantum associative memories that represent the ``quantization" of fully coupled neural networks like the Hopfield model. The idea is to replace the classical irreversible attractor dynamics driven by an Ising model with pattern-dependent weights by the reversible rotation of an input quantum state onto an output quantum state consisting of a linear superposition with probability amplitudes peaked on the stored pattern closest to the input in Hamming distance, resulting
in a high probability of measuring a memory pattern very similar to the input. The unitary operator implementing this transformation can be formulated as a sequence of one-qubit and two-qubit elementary quantum gates and is thus the exponential of an ordered quantum Ising model with sequential operations and with pattern-dependent interactions, exactly as in the classical case. Probabilistic quantum memories, that make use of postselection of the measurement result of control qubits, overcome the famed linear storage limitation of their classical counterparts because they permit to completely {\it eliminate crosstalk and spurious memories}. The number of control qubits plays the role of an inverse fictitious temperature, the accuracy of pattern retrieval can be tuned by lowering the fictitious temperature under a critical value for quantum content association while the complexity of the retrieval algorithm remains polynomial for any number of patterns polynomial in the number of qubits.  These models solve thus the capacity shortage problem of classical associative memories, providing a {\it polynomial improvement} in capacity. The price to pay is the probabilistic nature of information retrieval.

\end{abstract}
\maketitle

\section{Introduction}
There is a growing consensus that the fundamental mechanism of human intelligence is simply pattern recognition, the retrieval of information based on content association, albeit repeated in ever increasing hierarchical structures \cite{intelligence}. Correspondingly, pattern recognition in machine intelligence \cite{machineintel} has made enormous progress in the last decade or so and such systems are now to be found in applications ranging from medical diagnosis to facial and voice recognition in security and digital personal assistants, the latest addition to the the family being self-driving cars. On the other side, the last two decades have seen the birth of, and an explosion of research in a new information-theoretic field: quantum information theory and quantum computation \cite{review}. This chapter deals with quantum pattern recognition, with particular emphasis on models that are both accessible to detailed analytical treatment and efficiently implementable within the framework of the quantum circuit model. 

Pattern recognizers, which go also under the name of {\it associative memories} (or more precisely autoassociative memories), are fundamentally different than von Neumann or Turing machines \cite{turing}, which have grown into the ubiquitous computers that permeate our information society. Computation is not sequential but, rather, based on collective phenomena due to interactions among a large number of, typically redundant, elementary components. Information is not address-oriented, i.e. stored in look-up tables (random access memories, RAMs) but, rather, distributed in often very complex ways over the connections and interactions parameters. In traditional computers information is identified by a label and stored in a database indexed by these labels. Retrieval requires the exact knowledge of the relevant label, without which information is simply not accessible. This is definitely not how our own brain works. When trying to recognize a person from a blurred photo it is totally useless to know that it is the 16878th
person you met in your life. Rather, the recognition process is based on our strong power of association with stored memories that resemble the given picture. Association is what we use every time we solve a crossword puzzle and
is distinctive of the human brain.

The best known examples of pattern recognizers are neural networks \cite{neuralnetworks} and hidden Markov models \cite{markov}, the Hopfield model \cite{hopfield} (and its generalization to a bidirectional associative memory \cite{kosko}) being the paradigm, since it can be studied analytically in detail by the techniques of statistical mechanics \cite{neuralnetworks, statmech}. The great advantage of these architectures is that they eliminate the extreme rigidity of RAM memories, which require a precise knowledge of the memory address and, thus, do not permit the retrieval of incomplete or corrupted inputs. In associative memories, on the contrary, recall of information is possible also on the basis of partial knowledge of its content, without knowing a precise storage location, which typically does not even exist. This is why they are also called ``content-addressable memories". 

Unfortunately, classical associative memories suffer from  a severe capacity shortage. When storing multiple patterns, these interfere with each other, a phenomenon that goes under the name of crosstalk. Above a critical number of patterns, crosstalk becomes so strong that a phase transition to a completely disordered spin glass phase \cite{parisi} takes place. In this phase there is no relation whatsoever between the information encoded in the memory and the original patterns. For the Hopfield model, the critical threshold on the number $p$ of patterns that can be stored in a network of $n$ binary neurons is $p_{max} \simeq 0.138 \ n$ \cite{neuralnetworks} . While various possible improvements can be envisaged, the maximum number of patterns remains linear in the number of neurons, $p_{max} = O(n)$.

The power of quantum computation \cite{review} is mostly associated with the speed-up in computing time it can provide with respect to its classical counterpart, the paramount examples being Shor's factoring algorithm \cite{shor} and Grover's database search algorithm \cite{grover}. The efficiency advantage over classical computation is due essentially to the quantum superposition principle and entanglement, which allow for massively parallel information processing. 

The bulk of the research effort in quantum computation has focused on the ``quantization" of the classical sequential computer architecture, which has led to the quantum circuit model \cite{review}, in which information processing is realized by the sequential application of a universal set of elementary one- and two-qubit gates to typically highly entangled quantum states of many qubits. The computation is said to be efficient if the desired unitary evolution of the quantum state can be realized by the application of a polynomial number (in terms of the number of involved qubits) of these elementary quantum gates. 

However, the question immediately arises if quantum mechanics can be applied successfully also to the collective information processing paradigm typical of machine intelligence algorithms and, specifically, if there are advantages in doing so. While this research has trailed the development of the quantum circuit model, it is presently experiencing a flurry of increased interest, so much so that last year NASA and Google have teamed up to found the Quantum Artificial Intelligence Laboratory, entirely dedicated to develop and advance machine intelligence quantum algorithms. 

While speed has been the main focus of quantum computation, it can be shown that quantum mechanics also offers a way out from the impossibility of reconciling the association power of content-addressable memories with the requirement of large storage capacity. Indeed, one of us pointed out already in 2001 \cite{tru1, tru2, tru3} that storage capacity of associative memories can also be greatly enhanced by the quantum superposition principle. The key idea is to exploit the fundamental probabilistic nature of quantum mechanics. If one is willing to abandon the classical paradigm of one-off retrieval and sacrifice some speed by repeating the information retrieval step several times, then it is possible to store any desired polynomial number (in terms of the number of qubits) of patterns in a quantum associative memory and still tune the associative retrieval to a prescribed accuracy, a large advantage with respect to the classical linear limitation described above. Quantum entanglement permits to completely eliminate crosstalk and spurious memories in a tuneable probabilistic content association procedure with polynomial complexity for a polynomial number of stored patterns. Such probabilistic quantum associative memories can thus be implemented efficiently. Similar ideas in this direction were developed simultaneously in \cite{csj}. 

In this chapter we will review our own work on fundamental aspects of quantum associative memories and quantum pattern recognition. We will begin by a short survey of the main features of classical fully coupled neural networks like the Hopfield model and its generalizations, with a special emphasis on the capacity limitation and its origin. We will then describe the quantization of the Hopfield model \cite{dia}: the idea is to replace the classical irreversible dynamics that attracts input patterns to the closest minima of an energy function, representing the encoded memories, with a reversible unitary quantum evolution that amplifies an input quantum state to an output quantum state representing one of the stored memories at a given computational time $t$. In the classical model there is a complex phase diagram in terms of the two noise parameters, the temperature $T$ and the disorder $p/n$ with $n$ the number of bits and $p$ the number of stored patterns. It is, specifically the disorder due to an excessive loading factor $p/n$ that prevents the storage of more than a critical number of patterns by causing the transition to a spin glass phase \cite{parisi}, even at zero temperature. Correspondingly, in the quantum version there are quantum phase transitions due to both disorder and quantum fluctuations, the latter being encoded in the effective coupling $Jt$, with $J$ being the energy parameter of the model and $t$ being the computational time (throughout the review we will use units in which $c=1$ and $\hbar =1$). These are first examples of quantum collective phenomena typical of quantum machine intelligence. It turns out that, barring periodicity effects due to the unitary time evolution, the phase diagram for the quantum Hopfield model is not so different from its classical counterpart. Specifically, for small loading factors the quantum network has indeed associative power, a very interesting feature by itself, but the maximum loading factor is still limited to $p/n \le 1$, above which there is a totally disordered spin glass phase, with no association power for any computational time. The transition to this quantum spin glass phase takes place when one tries to store a number of memories that is not anymore linearly independent. 

We then turn our attention to probabilistic quantum associative memories \cite{tru1, tru2, tru3}. The basic idea underlying their architecture is essentially the same as above, with one crucial difference: they exploit, besides a unitary evolution, a second crucial aspect of quantum mechanics, namely wave function collapse upon measurement \cite{review}. A generic (pure) quantum state is a superposition of basis states with complex coefficients. A measurement projects (collapses) the state probabilistically onto one of the basis states, the probability distribution being governed by the squared absolute values of the superposition coefficients. Probabilistic quantum associative memories involve, besides the memory register itself a certain number $b$ of control qubits. The unitary evolution of the input state is again determined by a Hamiltonian that depends only on the stored patterns. Contrary to quantized Hopfield memories, however, this unitary evolution mixes the memory register and the control qubits. After having applied the unitary evolution to the initial input state , the control qubits are measured. Only if one obtains a certain specific result, one proceeds to measure the memory register. This procedure is called probabilistic postselection of the measurement result and guarantees that the memory register is in a superposition of the stored patterns such that the measurement probabilities are peaked on those patterns that minimize the Hamming distance to the input. A measurement of the memory register will thus associate input and stored patterns according to this probability distribution. 

Of course, if we limit ourselves to a maximum number $T$ of repetitions, there is a non-vanishing probability that the memory retrieval will fail entirely, since the correct control qubit state will never be measured.  One can say that information retrieval in these quantum memories consists of two steps: recognition (the correct state of the control qubits has been obtained) and identification (the memory register is measured to give an output). Both steps are probabilistic and both the recognition efficiency and the identification accuracy depend on the distribution of the stored patterns: recognition efficiency is best when the number of stored patterns is large and the input is similar to a substantial cluster of them, while identification accuracy is best for isolated patterns which are very different from all other ones, both very intuitive features. Both recognition efficiency and identification accuracy can be tuned to prescribed levels by varying the repetition threshold $T$ and the number $b$ of control qubits. 

The accuracy of the input-output association depends only on the choice of the number $b$ of control qubits. Indeed, we will show that  $t=1/b$ plays the role of an effective temperature \cite{tru3}. The lower $t$, the sharper is the corresponding effective Boltzmann distribution on the states closest in Hamming distance to the input and the better becomes the identification. By averaging over the distribution of stored patterns with Hamming distance to the input above a threshold $d$ one can eliminate the dependence on the stored pattern distribution and derive the effective statistical mechanics of quantum associative memories by introducing the usual thermodynamic potentials as a function of $d$ and the effective temperature $t=1/b$. 
In particular, the free energy $F(t)$ describes the average behaviour of the recall mechanism and provides concrete criteria to tune the accuracy of the quantum associative memory. By increasing $b$ (lowering $t$), the associative memory undergoes a phase
transition from a disordered phase with no correlation between input and output to an ordered phase with perfect input-output association encoded in the minimal Hamming distance $d$. This extends to quantum information theory the relation with Ising spin
systems known in error-correcting codes \cite{errcorr} and in public key cryptography \cite{crypto}.
 
The recognition efficiency can be tuned mainly by varying the repetition threshold $T$: the higher $T$, the larger the number of input qubits that can be corrupted without affecting recognition. The crucial point is that the recognition probability is bounded from below by $(p-1) (\pi/2)^{2b} / (p n^{2b})$. For any number of patterns, thus, a repetition threshold $T$ polynomial in $n$ guarantees recognition with probability $O(1)$. Due to the factor $(p-1)$ in the numerator, whose origin is exclusively quantum mechanical, the number of repetitions required for efficient recognition would actually be polynomial even for a number of patterns exponential in $n$. The overall complexity of probabilistic associative quantum memories is thus bounded by the complexity $O(p(2n+3))$ of the unitary evolution operator. Any polynomial number of patterns $p=O(n^x)$ can be encoded and retrieved efficiently in polynomial computing time. The absence of spurious memories leads to a substantial storage gain with respect to classical associative memories, the price to pay being the probabilistic nature of information recall.

\section{The classical Hopfield model}
Historically, the interest in neural networks \cite{neuralnetworks} has been driven by the desire to build machines capable of performing tasks for which the traditional sequential computer architecture is not well suited, like pattern recognition, categorization and generalization. Since these higher cognitive tasks are typical of biological intelligences, the design of these parallel distributed processing systems has been largely inspired by the physiology of the human brain.

The Hopfield model is one of the best studied and most successful neural networks. It was designed to model one particular higher cognitive function of the human brain, that of associative pattern retrieval or associative memory. 

The Hopfield model consists of an assembly of $n$ binary neurons $s_i$, $i=1 \dots n$ \cite{pitts}, which can take the values $\pm 1$ representing their firing (+1) and resting (-1) states. The neurons are fully connected by symmetric synapses with coupling strengths $w_{ij} = w_{ji}$ ($w_{ii} = 0$). Depending on the signs of these synaptic strengths, the couplings will be excitatory ($> 0$) or inhibitory ($< 0$). The model is characterised by an energy function
\begin{equation}
E=-{1\over 2} \sum_{i \ne j} w_{ij} \ s_i s_j \ , \ \ s_i = \pm 1 \ , \ \ i,j=1 \dots n \ ,
\label{bone}
\end{equation}
and its dynamical evolution is defined by the random sequential updating (in time $t$) of the neurons according to the rule
\begin{eqnarray}
s_i(t+1) &&= {\rm sign} \left( h_i(t) \right) \ ,
\label{btwoa}\\
h_i(t) &&= \sum_{i\ne j} w_{ij} s_j(t) \ ,
\label{btwob}
\end{eqnarray}
where $h_i$ is called the local magnetization. 

The synaptic coupling strengths are chosen according to the Hebb rule
\begin{equation}
w_{ij} = {1\over n} \sum_{\mu = 1 \dots p} \xi_i^{\mu} \xi_j^{\mu } \ ,
\label{bthree}
\end{equation}
where $\xi_i^{\mu }$, $\mu = 1 \dots p$ are $p$ binary patterns to be memorized. An associative memory is defined as a dynamical mechanism that, upon preparing the network in an initial state $s_i^0$ retrieves the stored pattern $\xi_i^{\lambda}$ that most closely resembles the presented pattern $s_i^0$, where resemblance is determined by minimizing the Hamming distance, i.e. the total number of different bits in the two patterns. As emerges clearly from this definition, all the memory information in a Hopfield neural network is encoded in the synaptic strengths.

It can be easily shown that the dynamical evolution (\ref{btwoa}) of the Hopfield model satisfies exactly the requirement for an associative memory. This is because:

\begin{itemize}
\item{} The dynamical evolution (\ref{btwoa}) minimizes the energy functional (\ref{bone}), i.e. this energy functional never increases when the network state is updated according to the evolution rule (\ref{btwoa}). Since the energy functional is bounded by below, this implies that the network dynamics must eventually reach a stationary point corresponding to a, possibly local, minimum of the energy functional.
\item{} The stored patterns $\xi_i^{\mu}$ correspond to, possibly local, minima of the energy functional. This implies that the stored patterns are attractors for the network dynamics (\ref{btwoa}). An initial pattern will evolve till it overlaps with the closest (in Hamming distance) stored pattern, after which it will not change anymore.
\end{itemize}

Actually, the second of these statements must be qualified. Indeed, the detailed behavior of the Hopfield model depends crucially upon the loading factor $\alpha = p/n$, the ratio between the number of stored memories and the number of available bits. This is best analyzed in the thermodynamic limit $p \to \infty$, $n \to \infty$, in which the different regimes can be studied by statistical mechanics techniques \cite{neuralnetworks, statmech} and characterized formally by the values of critical parameters.  

For $\alpha < \alpha1_{\rm crit} \simeq 0.051$, the system is in a ferromagnetic ($F$) phase in which there are global energy minima corresponding to all stored memories. The former differ from the original input memories only in a few percent of the total number of bits. Mixing between patterns leads to spurious local energy minima. These, however are destabilized at sufficiently high temperatures. 

For $\alpha1_{\rm crit} \simeq 0.051 < \alpha < \alpha2_{\rm crit} \simeq 0.138$ the system is in a mixed spin glass (SG) \cite{parisi} and ferromagnetic phase. There are still minima of sizable overlap with the original memories but they are now only metastable states. The true ground state is the spin glass, characterized by an exponentially increasing number of minima due to the mixing of original memories (crosstalk). The spin glass phase is orthogonal to all stored memories. If an input pattern is sufficiently near (in Hamming distance) to one of the original memories it will be trapped by the corresponding metastable state and the retrieval procedure is successful. On the other hand, if the input pattern is not sufficiently close to one of the stored memories, the network is confused and it will end up in a state very far from all original memories. 

For $\alpha > \alpha2_{\rm crit} \simeq 0.138$, the system is in a pure spin glass ($SG$) phase \cite{parisi} in which all retrieval capabilities are lost due to an uncontrolled proliferation of spurious memories. It is this phase transition to a spin glass that limits the storage capacity of the Hopfield model to $\alpha =p/n < 0.138$. While various improvements are possible, the storage capacity of classical associative memories remains linearly bounded by the number $n$ of classical bits \cite{neuralnetworks}. 

\section{Quantum neural networks and the quantization of the Hopfield model} 
In this section we introduce a quantum information processing paradigm that is different from the standard quantum circuit model \cite{dia}. Instead of one- and two-qubit gates that are switched on and off sequentially, we will consider long-range interactions that define a fully-connected quantum neural network of qubits. This is encoded in a Hamiltonian that generates a unitary evolution in which the operator acting on one qubit depends on the collective quantum state of all the other qubits. Note that some of the most promising technologies for the implementation of quantum information processing, like optical lattices \cite{optical} and arrays of quantum dots \cite{qudots} rely exactly on similar collective phenomena. 

In mathematical terms, the simplest classical neural network model is a graph with the following properties:
\begin{itemize}
\item{} A state variable $s_i$ is associated with each node (neuron) $i$.
\item{} A real-valued weight $w_{ij}$ is associated with each link (synapse) $(ij)$ between two nodes $i$ and $j$.
\item{} A state-space-valued transfer function $f(h_i)$ of the synaptic potential $h_i = \sum_j w_{ij} s_j$ determines the dynamics of the network.
\end{itemize}
Directed graphs correspond to feed-forward neural networks \cite{neuralnetworks} while undirected graphs with symmetric weights contain feed-back loops. If the graph is complete one has fully-connected neural networks like the Hopfield model. 
Two types of dynamical evolution have been considered: sequential or parallel synchronous. In the first case the
neurons are updated one at a time according to
\begin{equation}
s_i ( t + 1 ) = f \left( \sum_k w_{ik} s_k (t) \right) \ ,
\label{cone}
\end{equation}
while in the second case all neurons are updated at the same time. 
The simplest model is obtained when neurons become binary variables taking only the values $s_i=\pm 1$ for all $i$ and the transfer function becomes the sign function. This is the original McCullogh-Pitts \cite{pitts} neural network model, in 
which the two states represent quiescent and firing neurons. 

As we have seen in the previous section, the Hopfield model \cite{hopfield} is a fully-connected McCullogh-Pitts network 
in which the synaptic weights are symmetric quantities chosen according to the Hebb rule \cite{neuralnetworks}
\begin{equation}
w_{ij} = w_{ji} = {1\over n} \ \sum_{\mu =1}^p \xi_i^{\mu} \xi_j^{\mu} \ ,\qquad w_{ii}=0 \ .
\label{ctwo}
\end{equation}
and in which the the dynamics-defining function $f$ is the sign function, $f={\rm sign}$. This dynamics minimises the energy function
\begin{equation}
E=-{1\over 2} \sum_{i \ne j} w_{ij} \ s_i s_j \ , \ \ s_i = \pm 1 \ , \ \ i,j=1 \dots n \ ,
\label{cthree}
\end{equation}
where $n$ is the total number of neurons and ${\bf \xi}^{\mu}$ are the $p$ binary patterns to be memorized ($\xi_i^{\mu} = \pm 1$) 

A quantum McCullogh-Pitts network can correspondingly be defined as a graph that satisfies:
\begin{itemize}
\item{}A two-dimensional Hilbert space ${\cal H}_i$ is associated with each node (neuron) $i$, i.e. each neuron becomes a qubit whose basis states can be labeled as $|0>$ and $|1>$. 
\item{}A vector-valued weight $\vec{w}_{ij}$ is associated with each link (synapse) $(ij)$ between two nodes $i$ and $j$.
\item{}The synaptic potential becomes an operator $h_i = \sum_j \vec{w}_{ij} \vec{\sigma}_j$, where $\vec{\sigma}_i = \left(\sigma^x_i, \sigma^y_i, \sigma^z_i \right)$ is the vector of Pauli matrices acting on the Hilbert space ${\cal H}_i$. A unitary operator $U(h_i)$ determines the dynamics of the network starting from an initial input quantum state on the product Hilbert space of all qubits. 
\end{itemize}
In case of feed-forward quantum networks on directed graphs only a subset of qubits is measured after the unitary evolution, in case of fully connected quantum networks with symmetric weights the state of the whole network is relevant. 

The crucial difference with respect to classical neural networks concerns the interactions between qubits. In the classical model,  
the dynamics (\ref{cone}) induced by the transfer function is fully deterministic and irreversible, which is not compatible with quantum mechanics. A first generalization that has been considered is that of stochastic neurons, in which the transfer function determines only the probabilities that the classical state variables will take one of the two values: $n_i ( t + 1 ) = \pm 1$ with probabilities $f(\pm h_i(t))$, where $f$ must satisfy $f(h \to -\infty) = 0$, $f(h \to +\infty) = 1$ and $f(h)+f(-h)=1$. While this modification makes the dynamics probabilistic by introducing thermal noise, the evolution of the network is still irreversible since the 
actual {\it values} of the neurons are prescribed after an update step. In quantum mechanics the evolution must be reversible and only the magnitudes of the {\it changes} in the neuron variables can be postulated. Actually, the dynamics must generate a {\it unitary} evolution of the network. 

It is known that two-level unitary gates are universal, i.e. every unitary matrix on an $n$-dimensional Hilbert space may be written as a product of two-level unitary matrices. However, an arbitrary unitary evolution cannot be implemented as a sequential succession of a discrete set of elementary gates, nor can it be approximated efficiently with a polynomial number of such gates \cite{review}. In general, quantum neural networks as defined above, have to be thought of as defined by Hamiltonians $H$ that code hard-wired qubit interactions and generate a unitary evolution $U={\rm exp}(iHt)$. This corresponds to the parallel synchronous dynamics of classical neural networks. Only in particular cases, one of which will be the subject of the next section, does this unitary evolution admit a representation as a sequential succession of a discrete set of elementary one- and two-bit gates. In this cases the network admits a sequential dynamics as its classical counterpart. 

We now describe a direct ``quantization" of the Hopfield model in this spirit, i.e. by defining a quantum Hamiltonian that generalizes (\ref{cthree}). At first sight one would be tempted to simply replace the classical spins $s_i$ of (\ref{cthree}) with the third Pauli matrix $\sigma^z_i$ acting on the Hilbert space ${\cal H}_i$. This however would accomplish nothing, the model would still be identical to the original classical model, since all terms in the Hamiltonian would commute between themselves. A truly quantum model must involve at least two of the three Pauli matrices. In \cite{dia} we have proposed the following ``transverse" Hamiltonian:
\begin{equation}
{\cal H} = J \sum_{ij}\ w_{ij} \sigma^y_i \sigma^z_j \ ,
\label{cfour}
\end{equation}
where $\sigma ^k$, $k=x,y,z$ denote the Pauli matrices and $J$ is a coupling constant with the dimensions of mass (we remind the reader that we use units in which $c=1, \hbar =1$).  This generates a unitary evolution of the network:
\begin{equation}
|\psi (t)> = {\rm exp} (i{\cal H}t) \ |\psi_0> \ ,
\label{cfive}
\end{equation}
where $|\psi_0> = |\psi (t=0)>$. Specifically,
we will choose as initial configuration of the network the uniform 
superposition of all computational basis states \cite{review}
\begin{equation}
|\psi _0> = {1\over \sqrt{2^n}}\ \sum_{x=0}^{2^n-1} |x> \ .
\label{csix}
\end{equation} 
This corresponds to a "blank memory" in the sense that all possible states have the same probability of being recovered upon measurement. In the language of spin systems this is a state in which all spins are aligned in the $x$ direction.  

Inputs $\xi^{\rm ext}$ can be accomodated by adding an external transverse magnetic field along the $y$ axis, 
i.e. modifying the Hamiltonian to 
\begin{equation}
{\cal H}= J \sum_{ij} w_{ij} \sigma^y_i \sigma^z_j + g \sum_{i} h_i^{\rm ext} \sigma^y_i \ , 
\label{cseven}
\end{equation}
where $h_i^{\rm ext} = \sum_{j} w_{ij} \xi_j^{\rm ext}$. 
This external magnetic field
can be thought of as arising from the interaction of the network with an additional ``sensory" qubit register
prepared in the state $\xi^{\rm ext}$, the synaptic weights between the two layers 
being identical to those of the network self-couplings.  

Let us now specialize to the simplest case of one assigned memory ${\bf \xi}$ in which $w_{ij} = \xi_i \xi_j / n$. In the classical Hopfield model there are two nominal stable states that represent attractors for the dynamics, the pattern ${\bf \xi}$ itself 
and its negative $-{\bf \xi}$. Correspondingly, the quantum dynamics defined by the Hamiltonian (\ref{cfour})
and the initial state (\ref{csix}) have a $Z_2$ symmetry generated by $\prod_{i} \sigma^x_i$, corresponding to the 
inversion $|0> \leftrightarrow |1>$ of all qubits.  

As in the classical case we shall analyze the model in the mean field approximation. In this case, the mean field represents the average over quantum fluctuations rather than thermal ones but the principle remains the same. The mean field model becomes exactly solvable and allows to derive self-consistency conditions on the average overlaps with the stored patterns. In the classical case, the mean field approximation is known to become exact for long-range interactions \cite{stat}. 

In the quantum mean-field approximation operators are decomposed in a sum of their mean values in a given quantum state
and fluctuations around it, $\sigma_i^k = <\sigma_i^k> + \left( \sigma_i^k - <\sigma_i^k> \right)$, and quadratic
terms in the fluctuations are neglected in the Hamiltonian. Apart from an irrelevant constant, this gives
\begin{eqnarray}
{\cal H}_{\rm mf} &&= J \sum_i \sigma^y_i \left( <h^z_i> + {g\over J} h_i^{\rm ext}\right) + \sigma^z_i <h^y_i> \ ,
\nonumber \\
< h^k_i > &&= \sum_j w_{ij} <\sigma^k_j> = \xi_i \ m^k \ ,
\label{ceight}
\end{eqnarray}
where $m^k = (1/n)\ \sum_i <\sigma^k_i> \xi_i$ is the average overlap of the state of the network with the stored pattern.
This means that each qubit $i$ interacts with the average
magnetic field (synaptic potential) $< h^k_i >$ due to all other qubits: naturally, the correct values of these 
mean magnetic fields $<h^k_i>$ have to be determined self-consistently. 

To this end we compute the average pattern overlaps $m^k$ using the mean field Hamiltonian (\ref{ceight}) to
generate the time evolution of the quantum state. This reduces to a sequence of factorized rotations
in the Hilbert spaces of each qubit, giving
\begin{eqnarray}
m^y &&= -{m^y \over \vert m\vert} \ {\rm sin} \ 2Jt \vert m\vert \ ,
\nonumber \\
m^z &&= {{m^z + (g/J)M^z} \over \vert m\vert} \ {\rm sin} \ 2Jt \vert m\vert \ ,
\label{cnine}
\end{eqnarray}
where $\vert m\vert = \sqrt{\left( m^y \right)^2 + \left( m^z + (g/J)M^z \right)^2}$ and 
$M^z = (1/n) \ \sum_i \xi_i^{\rm ext} \xi_i$ is the average overlap of the external stimulus with the stored memory.

Before we present the detailed solution of these equations, 
let us illustrate the mechanism underlying the quantum associative memory. 
To this end we note that, for $g=0$, the pattern overlaps $m^y$ and $m^z$ in the two directions cannot
be simultaneously different from zero. As we show below, only $m^z \ne 0$ for $J>0$ (for $J<0$ the
roles of $m^y$ and $m^z$ are interchanged). In this case 
the evolution of the network becomes a sequence of $n$ rotations
\begin{equation}
\left( \begin{array}{cc}
{\rm cos}(Jt <h^z_i>) & {\rm sin}(Jt <h^z_i>) \\
-{\rm sin}(Jt <h^z_i>) & {\rm cos}(Jt <h^z_i>) \\
\end{array} \right)
\label{cten}
\end{equation}
in the two-dimensional Hilbert spaces of each qubit $i$. The rotation parameter is exactly the same synaptic potential
$h_i$ which governs the classical dynamics of the Hopfield model.  
When these rotations are applied on the initial state (\ref{csix}) they amount to a single update step
transforming the qubit spinors into 
\begin{equation}
{1\over \sqrt{2}} \ \left( \begin{array}{cc}
{\rm cos}(Jt <h^z_i>) + {\rm sin}(Jt <h^z_i>) \\
{\rm cos}(Jt <h^z_i>) - {\rm sin}(Jt <h^z_i>)\\
\end{array} \right) \ .
\label{celeven}
\end{equation}
This is the generalization to quantum probability {\it amplitudes} of the probabilistic formulation of classical stochastic neurons.
Indeed, the probabilities for the qubit to be in its eigenstates $\pm 1$ after a time $t$, obtained by squaring the probability amplitudes, are given 
by $f(\pm <h^z>)$, where $f(<h^z>) = (1+{\rm sin}(2Jt <h^z>))/2$ has exactly 
the properties of an activation function (alternative to the Fermi function), at
least in the region $Jt < \pi/4$. In this correspondence, the effective coupling constant $Jt$ plays
the role of the inverse temperature, as usual in quantum mechanics. 

We shall now focus on a network without external inputs. In this case the equation for the average pattern overlaps has
only the solution $\vert m\vert=0$ for $0< Jt < 1/2$.  
For such small effective couplings (high effective temperatures), corresponding to weak synaptic connections or to
short evolution times, the network is unable to remember the
stored pattern. For $1/2 < Jt $, however, the solution $\vert m\vert =0$ becomes unstable, 
and two new stable solutions $m^z=\pm m_0$ appear. 
This means that the reaction of the
mean orientation of the qubit spinors against a small deviation $\delta m^z$ from the $\vert m\vert =0$ solution is larger than the deviation itself. 
Indeed, any so small external perturbation $(g/J)M^z$ present at the bifurcation time $t=1/2J$ is sufficient for the network evolution
to choose one of the two stable solutions, according to the sign of the external perturbation. The point $Jt =1/2$ 
represents a quantum phase transition \cite{sachdev} from an amnesia (paramagnetic) phase
to an ordered (ferromagnetic) phase in which the network has recall capabilities: the average pattern overlap $m^z$ is the corresponding order
parameter. In the ferromagnetic phase the original $Z_2$ symmetry of the model is spontaneously broken. 

%\begin{figure}
%\includegraphics[width=8cm]{qh1.eps}
%\caption{\label{fig:Fig1} The order parameter of quantum associative memories.}
%\end{figure}

For $Jt = \pi/4 $, the solution becomes $\vert m_0\vert =1$, which means that the network is capable of perfect recall
of the stored memory.
For $Jt > \pi/4$ the solution $m_0$ decreases slowly to 0 again. Due to the periodicity of the time evolution, however,
new stable solutions $m_0 = \pm 1$ appear at $Jt = (1+4n)\pi/4$ for every integer $n$. Also, for $Jt \ge 3\pi/4$, new solutions
with $m^y \ne 0$ and $m^z=0$ appear. These, however, correspond all to metastable states.
Thus, $t=\pi/4J$ is the ideal computation time for the network. 

The following picture of quantum associative memories emerges from the above construction. States of
the network are generic linear superpositions of computational basis states. The network is prepared in the state $|\psi _0>$ and
is then let to unitarily evolve for a time $t$. After this time the state of the network is measured, giving the result
of the computation. During the evolution each qubit updates its quantum state by a rotation that depends on the aggregated
synaptic potential determined by the state of all other qubits. These synaptic potentials are subject to large quantum fluctutations
which are symmetric around the mean value $<h^z>=0$. If the interaction is strong enough, any external disturbance will cause the 
fluctuations to collapse onto a collective rotation of all the network's qubits towards the nearest memory. 
 
We will now turn to the more interesting case of a finite density $\alpha =p/n$ of stored memories in the limit $n\to \infty$.
In this case the state of the network can have a finite overlap with several stored memories $\xi ^{\mu}$ simultaneously. As in the
classical case we shall focus on the most interesting case of a single "condensed pattern", in which the network uniquely recalls one
memory without admixtures.  
Without loss of generality we will chose this memory to be the first, $\mu =1$, omitting then the memory superscript on the
corresponding overlap $m$. Correspondingly we will consider external inputs so that only $M^{\mu=1} =M\ne 0$. 
For simplicity of presentation, we
will focus directly on solutions with a non-vanishing pattern overlap along the z-axis, omitting also the direction superscript $z$. 

In case of a finite density of stored patterns, one cannot neglect the noise effect due to the infinite number of memories. 
This changes (\ref{cnine}) to
\begin{eqnarray}
m &&= {1\over n} \sum_i \ {\rm sin} \ 2Jt\left( m + {g\over J} M + \Delta_i\right) \ ,
\nonumber \\
\Delta_i &&= \sum_{\mu \ne 1} \xi_i^1 \xi_i^{\mu} m^{\mu} \ .
\label{ctwelve}
\end{eqnarray} 
As in the classical case we will assume that $\{ \xi_i^{\mu}\}$ and $\{ m^{\mu}, \mu \ne 1\}$ are all independent random variables
with mean zero and we will denote by square brackets the configurational average over the distributions of these random variables.
As a consequence of this assumption, the mean and variance of the noise term are given by $[\Delta_i] = 0$ and
$[\Delta_i^2] = \alpha r$, where 
\begin{equation}
r = {1\over \alpha} \ \sum_{\mu \ne 1} \left[ \left( m^{\mu} \right)^2 \right]
\label{cthirteen}
\end{equation}
is the spin-glass order parameter \cite{parisi}. According to the central limit theorem one can now
replace $n^{-1} \sum_i$ in (\ref{ctwelve}) by an average over a Gaussian noise,
\begin{equation}
m = \int {dz\over \sqrt{2\pi}} \ {\rm e}^{-z^2 \over 2} \ {\rm sin} \ 2Jt \left( m + {g\over J} M + \sqrt{\alpha r} z \right) \ .
\label{cfourteen}
\end{equation}

The second order parameter $r$ has to be evaluated self-consistently by a similar procedure starting from the equation
analogous to eq. (\ref{ctwelve}) for $\mu \ne 1$. In this case one can use $m^{\mu } \ll 1$ for $\mu \ne 1$ to expand the transcendental function
on the right-hand side in powers of this small parameter, which gives
\begin{eqnarray}
v &&= \int {dz\over \sqrt{2\pi}} \ {\rm e}^{-z^2 \over 2} 
\ {\rm sin}^2 \ 2Jt \left( m + {g\over J} M + \sqrt{\alpha r} z \right) \ ,
\nonumber \\
x &&= \int {dz\over \sqrt{2\pi}} \ {\rm e}^{-z^2 \over 2} \ {\rm cos} \ 2Jt \left( m + {g\over J} M + \sqrt{\alpha r} z \right) \ ,
\label{cfifteen}
\end{eqnarray}
where $v = (1-2Jtx)^2 r$. 
Solving the integrals gives finally the following coupled equations for the two order parameters $m$ and $r$:
\begin{eqnarray}
m &&= {\rm sin} \ 2Jt \left( m+{g\over J} M \right) \ {\rm e}^{-2 (Jt)^2 \alpha r} \ ,
\nonumber \\
r &&= {1\over 2} \ {{1-{\rm cos} \ 4Jt \left( m+{g\over J} M\right) \ {\rm e}^{-8 (Jt)^2 \alpha r}} \over
{\left( 1-2Jt {\rm cos} \ 2Jt \left( m+{g\over J} M\right) \ {\rm e}^{-2 (Jt)^2 \alpha r} \right)^2}} \ .
\label{csixteen}
\end{eqnarray}

In terms of these order parameters one can distinguish three phases of the network. First of all the value of $m$ determines
the presence ($m>0$) or absence ($m=0$) of ferromagnetic order (F). If $m=0$ the network can be in a paramagnetic phase (P) if also $r=0$ or a quantum spin glass phase (SG) if $r>0$. The phase structure resulting from a numerical solution of the coupled equations (\ref{csixteen}) for $g=0$ is shown in Fig. 1. 

\begin{figure}
\includegraphics[width=14cm]{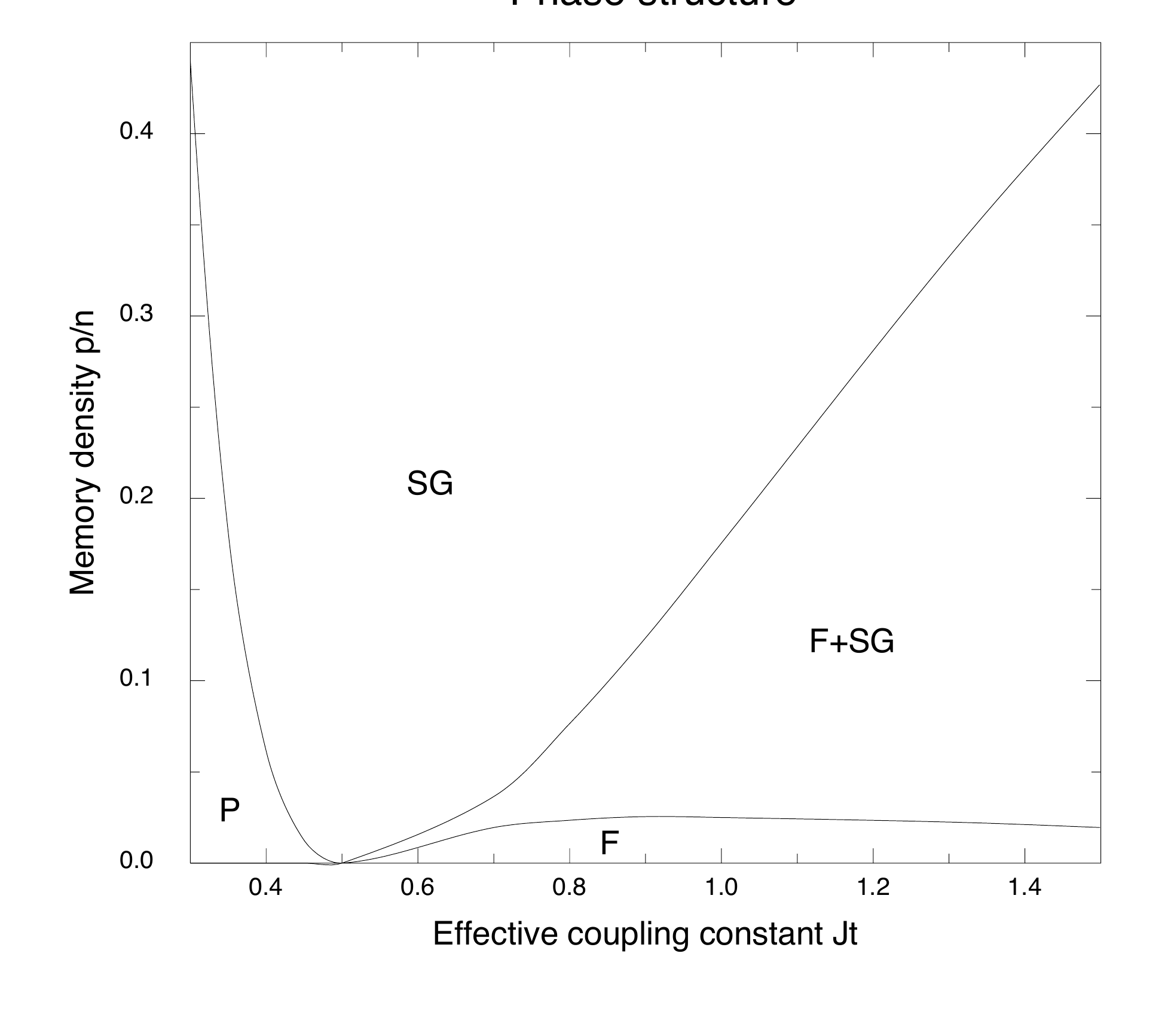}
\caption{\label{fig:Fig1} The phase structure of quantum associative memories with finite density of stored patterns. P, F and SG 
denote (quantum) paramagnetic, ferromagnetic and spin-glass phases, respectively. F + SG denotes a mixed phase in which the 
memory retrieval solution is only locally stable.}
\end{figure}

For $\alpha < 0.025$ the picture is not very different from the single memory case. For large enough computation times
there exists a ferromagnetic phase in which the $m=0$ solution is unstable and the network has recall capabilities. The only
difference is that the maximum value of the order parameter $m$ is smaller than 1 (recall is not perfect due to noise) 
and the ideal computation time $t$ at which the maximum is reached depends on $\alpha$. For $0.025 <\alpha <1.000$
instead, ferromagnetic order coexists as a metastable state with a quantum spin glass state. This means that ending up in the 
memory retrieval solution depends not only on the presence of an external stimulus but also on its magnitude; in other words, the
external pattern has to be close enough to the stored memory in order to be retrieved. 
For $1<\alpha $ all retrieval capabilities are lost and the network will be in a quantum spin glass state for all computation
times (after the transition from the quantum paramagnet). $\alpha =1$ is thus the maximum memory capacity of this quantum network.
Note that $\alpha = 1$ corresponds to the maximum possible number of linearly independent memories. 
For memory densities smaller but close to this maximum value, however,
the ferromagnetic solution exists only for a small range of effective couplings centered around
$Jt \simeq 9$: for these high values of $Jt$ the quality of pattern retrieval is poor, the value of the order parameter $m$ being
of the order 0.15-0.2. Much better retrieval qualities are obtained for 
smaller effective couplings: e.g. for $Jt=1$ the order parameter is larger than 0.9 (corresponding to an error rate smaller than 5\%) for memory densities up to 0.1. In this case, however the maximum memory density is 0.175, comparable with the classical result of the Hopfield model. Quantum mechanics, here, does not carry any advantage. 

\section{Probabilistic Quantum Memories}
We have seen in the last section that crosstalk prevents the amplification of patterns stored in the weights of a simple quantum Hamiltonian like (\ref{cfour}) when the loading factor exceeds a linear bound comparable with the classical one. In this section we show that this limit can be overcome by probabilistic quantum memories, which use postselection of the measurement results of certain control qubits \cite{tru1, tru2, tru3}. The price to pay is that such probabilistic memories require repetitions of the retrieval process and that there is non-vanishing probability that this fails entirely. When it is successful, however, it allows retrieval of the most appropriate pattern among a polynomial pool instead of a linear one. 

\subsection{Storing patterns}
Let us start by describing the elementary quantum gates \cite{review} that we will use in the rest of the paper. First of all there are the single-qbit gates represented by the Pauli matrices $\sigma^i$, $i=x,y,z$. The first Pauli matrix $\sigma^x$, in particular, implements the NOT gate. Another single-qbit gate is the Hadamard gate H, with the matrix representation
\begin{equation}
H = {1\over \sqrt{2}} \ \left( \begin{matrix} 1 & 1\\
1 & -1\\\end{matrix} \right) \ .
\label{adda}
\end{equation}
Then, we will use extensively the two-qbit XOR (exclusive OR) gate, which performs a NOT on the second qbit if and only if the first one is in state $|1\rangle$. In matrix notation this gate is represented as ${\rm XOR} = {\rm diag} \left( 1, \sigma^x \right)$, where $1$ denotes a two-dimensional identity matrix and $\sigma^x$ acts on the components $|01\rangle$ and $|11\rangle$ of the Hilbert space. The 2XOR, or Toffoli gate is the three qbit generalization of the XOR gate: it performs a NOT on the third qbit if and only if the first two are both in state $|1\rangle$. In matrix notation it is given by ${\rm 2XOR} = {\rm diag} \left( 1, 1, \sigma^x \right)$. In the storage algorithm we shall make use also of the nXOR generalization of these gates, in which there are n control qbits. This gate is also used in the subroutines implementing the oracles underlying Grover's algorithm \cite{review} and can be realized using unitary maps affecting only few qbits at a time \cite{gates}, which makes it efficient. All these are standard gates. In addition to them we introduce the two-qbit controlled gates
\begin{eqnarray}
CS^i &&= |0\rangle \langle 0| \otimes 1 + |1\rangle \langle 1|
\otimes S^i \ ,
\nonumber \\
S^i &&= \left( \begin{matrix}\sqrt{i-1\over i}&1\over \sqrt{i}\\
-1\over{\sqrt{i}}&\sqrt{i-1\over i}\\ \end{matrix} \right) \ ,
\label{addb}
\end{eqnarray}
for $i=1, \dots, p$. These have the matrix notation $CS^i = {\rm diag} \left( 1, S^i \right)$. For all these gates we shall indicate by subscripts the qbits on which they are applied, the control qbits coming always first.  

The construction of quantum memories relies, of course, on the fundamental fact that one can use entanglement to "store" an arbitrary number $p$ of binary patterns $p^i$ of length $n$ in a quantum superposition of just $n$ qubits, 
\begin{equation}
|m\rangle = {1\over \sqrt{p}}\ \sum_{i=1}^p \ |p^i\rangle \ .
\label{a}
\end{equation}
The idea of the memory architecture consists thus of two steps:
\begin{itemize}
\item{} Generate the state $|m\rangle$ by a unitary evolution $M$ from a simple prepared state, say $|0_1, \dots, 0_n\rangle$, 
$|m\rangle = M |0_1, \dots, 0_n\rangle $. 
\item{} Given an input state $|i\rangle = |i_1,\dots, i_n\rangle$, generate from $|m\rangle$ a superposition of the pattern states that is no more uniform but whose amplitudes define a probability distribution peaked on the pattern states with minimal Hamming distance front the input. It is this step that involves both a unitary evolution and a postselection of the measurement result. 
\end{itemize}
The quantum memory itself is the unitary operator $M$ that codes the $p$ patterns. It defines implicitly a Hamiltonian through the formal relation $M= {\rm exp} (i {\cal H})$, a Hamiltonian that represents pattern-dependent interactions among the qubits. This is the quantum generalization of the classical Hopfield model. In order to dispel any possible misunderstandings right away, we point out that this is quite different to the communication of classical information via a quantum channel, limited by the Holevo theorem \cite{holevo}, as we discuss in detail below. 

In order to construct explicitly the quantum memory $M$ we will start from an algorithm that loads sequentially the classical patterns into an auxiliary register, from which they are then copied into the actual memory register. A first version of such an algorithm was introduced in \cite{ventura}. The simplified version that we present here is due to \cite{tru1}. 

We shall use three registers: a first register $p$ of $n$ qbits in which we will subsequently feed the patterns $p^i$ to be stored, a
utility register $u$ of two qbits prepared in state $|01\rangle$, and another register $m$ of $n$ qbits to hold the memory. This latter will be initially prepared in state$|0_1, \dots, 0_n\rangle$. The full initial quantum state is thus
\begin{equation}
|\psi _0^1\rangle = |p^1_1, \dots p^1_n; 01; 0_1, \dots, 0_n\rangle \ .
\label{c}
\end{equation}
The idea of the storage algorithm is to separate this state into two terms, one corresponding to the already stored patterns, and another ready to process a new pattern. These two parts will be distinguished by the state of the second utility qbit $u_2$: $|0\rangle$ for the stored patterns and $|1\rangle$ for the processing term. 

For each pattern $p^i$ to be stored one has to perform the operations described below:
\begin{equation}
|\psi _1^i \rangle = \prod_{j=1}^n 
\ 2XOR_{p_j^i u_2 m_j} \ |\psi _0^i\rangle \ .
\label{addc}
\end{equation}
This simply copies pattern $p^i$ into the memory register of the processing term, identified by $|u_2\rangle = |1\rangle$.
\begin{eqnarray}
|\psi _2^i \rangle &&= \prod_{j=1}^n 
\ NOT_{m_j} \ XOR_{p_j^i m_j} \ |\psi _1^i\rangle \ ,
\nonumber \\
|\psi _3^i \rangle &&= nXOR_{m_1 \dots m_n u_1} |\psi_2^i \rangle \ .
\label{addd}
\end{eqnarray}
The first of these operations makes all qbits of the memory register $|1\rangle$'s when the contents of the pattern and memory registers are identical, which is exactly the case only for the processing term. Together, these two operations change the first utility qbit $u_1$ of the processing term to a $|1\rangle$, leaving it unchanged for the stored patterns term.
\begin{equation}
|\psi_4^i\rangle = CS^{p+1-i}_{u_1 u_2} \ |\psi _3^i\rangle \ .
\label{adde}
\end{equation}
This is the central operation of the storing algorithm. It separates out the new pattern to be stored, already with the correct normalization factor.
\begin{eqnarray}
|\psi _5^i \rangle &&= nXOR_{m_1 \dots m_n u_1} |\psi_4^i \rangle \ ,
\nonumber \\
|\psi _6^i \rangle &&= \prod_{j=n}^1
\ XOR_{p_j^i m_j} \ NOT_{m_j}\  |\psi _5^i\rangle \ .
\label{addf}
\end{eqnarray}
These two operations are the inverse of eqs.(\ref{addd}) and restore the utility qbit $u_1$ and the memory register $m$ to their original values. After these operations on has
\begin{equation}
|\psi _6^i \rangle = 
{1\over \sqrt{p}}\ \sum_{k=1}^i |p^i;00;p^k\rangle + \sqrt{p-i\over p}
|p^i;01;p^i\rangle \ .
\label{d}
\end{equation}
With the last operation,
\begin{equation}
|\psi_7^i\rangle = \prod_{j=n}^1\ 2XOR_{p^i_j u_2 m_j}
\ |\psi_6^i\rangle \ ,
\label{addg}
\end{equation}
one restores the third register $m$ of the processing term, the second term in eq.(\ref{d}) above, to its initial value $|0_1, \dots, 0_n\rangle$. At this point one can load a new pattern into register $p$ and go through the same routine as just described. At the end of the whole process, the $m$-register is exactly in state $|m\rangle$, eq. (\ref{a}).

Any quantum state can be generically obtained by a unitary transformation of the initial state $|0,\dots ,0\rangle$. This is true also for the memory state $|m\rangle$. In the following we will explicitly construct the unitary memory operator $M$ which implements the transformation $|m\rangle = M \ |0,\dots ,0\rangle$.

To this end we introduce first the single-qbit unitary gates 
\begin{equation}
U^i_j = {\rm cos} \left( {\pi \over 2} p^i_j \right) 1 + i \ {\rm sin} \left(
{\pi \over 2} p^i_j \right) \sigma _2 \ ,
\label{newa}
\end{equation}
where $\sigma_2$ is the second Pauli matrix. These operators are such that their product over the $n$ qbits generates pattern $p^i$ out of $|0,\dots, 0\rangle$:
\begin{eqnarray}
|p^i\rangle &&= P^i \ |0,\dots ,0\rangle \ ,
\nonumber \\
P^i &&\equiv \prod_{j=1}^n U^i_j \ .
\label{newb}
\end{eqnarray}
We now introduce, in addition to the memory register proper, the same two utility qbits as before, also initially in the state $|0\rangle$. The idea is, exactly as in the sequential algorithm, to split the state into two parts, a storage term with $|u_2\rangle = |0\rangle$ and a processing term with $|u_2\rangle = |1\rangle$. Therefore we generalize the operators $P^i$ defined
above to
\begin{equation}
CP^i_{u_2} \equiv \prod_{j=1}^n \ CU^i_{u_2 j} \ ,
\label{newc}
\end{equation}
which loads pattern $p^i$ into the memory register only for the processing term. It is then easy to check that
\begin{eqnarray}
&&|m;00\rangle = M \ |0,\dots, 0;00\rangle \ ,
\nonumber \\
&&M = \prod_{i=1}^p \left[ \left( CP^i_{u_2} \right) ^{-1} NOT_{u_1}
CS^{p+1-i}_{u_1 u_2} XOR_{u_2 u_1} CP^i_{u_2} \right] \times 
\nonumber \\
&&\times \ NOT_{u_2} \ .
\label{newd}
\end{eqnarray}
From this construction it is easy to see that the memory operator $M$ involves a number $p(2n+3) + 1$ of elementary one- and two-qbit gates. It is thus efficient for any number $p$ of patterns polynomial in the number $n$ of qubits. It is interesting to note that another version of this operator has been recently derived in \cite{tanaka}, with a bound of $O(pn^3/6)$ on its complexity. This is also linear in $p$, implying  again efficiency for a polynomial number of patterns. 

While the memory construction we have presented here mirrors its classical counterpart, it is important to stress one notable difference. In classical associative memories, patterns are stored as minima of an energy landscape or, alternatively in the parameters of a dynamical evolution law \cite{neuralnetworks}. This is reflected verbatim in the construction of the unitary operator $M$ in (\ref{newd}), which completely codes the patterns in a dynamical law, albeit reversible in the quantum case. In quantum mechanics, however, there is the possibility of shuffling some (but not all, as we will shortly see) information about the patterns from the unitary evolution law $M$ onto a set of quantum states. 

The ideal, most compressed quantum memory would indeed be the quantum superposition of patterns $|m\rangle$ in (\ref{a}) itself. This, however is impossible. If the memory state has to be used for information retrieval it must be measured and this destroys all information about the patterns (save the one obtained in the measurement). The quantum state must therefore be copied prior to use and this is impossible since the linearity of quantum mechanics forbids exact universal cloning of quantum states \cite{zurek}. Universal cloning of quantum states is possible only in an approximate sense \cite{unclo} and has two disadvantages: first of all the copies are imperfect, though optimal \cite{opt} and secondly, the quality of the master copy decreases with each additional copy made. Approximate universal cloning is thus excluded for the purposes of information recall since the memory would be quickly washed out. 

This leaves state-dependent cloning \cite{sdclo} as the only viable option. State-dependent cloners are designed to reproduce only a finite number of states and this is definitely enough for our purposes. Actually the memory $M$ in (\ref{newd}) is equivalent to a state-dependent cloner for the the state $|m\rangle$ in (\ref{a}). In this case the information about the stored patterns is completely coded in the memory operator, or equivalently, the state-dependent cloner. It is possible, however, to subdivide the pattern information among an operator and a set of quantum states, obviously including $|m\rangle$, by using a probabilistic cloning machine \cite{cloning}. Probabilistic cloners copy quantum states exactly but the copying process is not guaranteed to succeed and must be repeated until the measurement of an auxiliary register produces a given result associated with copying success. In general, any number of linearly independent states can be copied probabilistically. In the present case for example, it would be sufficient to consider any dummy state $|d\rangle$ different from $|m\rangle$ (for more than two states the condition would be linear independence) and to construct a probabilistic cloning machine for these two states. This machine would reproduce
$|m\rangle$ with probability $p_m$ and $|d\rangle$ with probability $p_d$; a flag would tell when the desired state $|m\rangle$ has been obtained. In order to obtain an exact copy of $|m\rangle$ one would need then $1/p_m$ trials on average. The master copy would be exactly preserved. 

The cloning efficiencies of the probabilistic cloner of two states are bounded as follows \cite{cloning}:
\begin{equation}
p_m + p_d \le {2\over 1+ \langle d|m \rangle } \ .
\label{neweqa}
\end{equation}
This bound can be made large by choosing $|d\rangle$ as nearly orthogonal to $|m\rangle$ as possible. A simple way to achieve this for a large number of patterns would be, for example, to encode also the state
\begin{equation}
|d\rangle = {1\over \sqrt{p}} \ \sum_{i=1}^p (-1)^{i+1} |p^i \rangle 
\label{neweqb}
\end{equation}
together with $|m\rangle$ when storing information. This can be done easily by using alternately the operators $S^i$ and $\left( S^i \right) ^{-1}$ in the storing algorithm above. For binary patterns which are all different from one would then have
\begin{eqnarray}
\langle d|m\rangle &&= 0 \ , \qquad \qquad p \ {\rm even} \ ,\\
\nonumber
\langle d|m\rangle &&= {1\over p} \ , \qquad \qquad p \ {\rm odd} \ ,
\label{neweqc}
\end{eqnarray}
and the bound for the cloning efficiencies would be very close to its maximal value 2 in both cases. 

The quantum network for the probabilistic cloner of two states has been developed in \cite{clonet}. It can be constructed exclusively
out of the two simple distinguishability tranfer (D) and state separation (S) gates. As expected, these gates embody information about the two states to be cloned. Part of the memory, therefore, still resides in the cloning network. The pattern-dependence of the network cloner can be decreased by choosing a larger set of states in the pool that can be cloned, so that the cloner becomes more and more generic. On one side this decreases also the efficiency of the cloner, so that more repetitions are required, on the other side, since the clonable pool is limited to a set of linearly independent states, one can never eliminate completely the pattern-dependence of the cloning operator. This is why the original claim of an exponential capacity increase of quantum associative memories \cite{tru1}, based on probabilistic cloning of the state $|m\rangle$, is excessive. The complexity of the cloner, be it exact as in the memory operator $M$ or probabilistic, remains linear in the number of patterns and the requirement of efficient implementability limits thus $p$ to a polynomial function of the number $n$ of qubits., which is still a large improvement upon classical associative memories. 

\subsection{Retrieving patterns}
Let us now assume we are given a binary input $i$ that is a corrupted version of one of the patterns stored in the memory. The task of the retrieval algorithm is to "recognize" it, i.e. output the stored pattern that most resembles this input, where similarity is defined (here) in terms of the Hamming distance, the number of different bits between the two patterns, although other similarity measures \cite{nerualnetworks} could also be incorporated. 

The retrieval algorithm requires also three registers. The first register $i$ of n qbits contains the input pattern; 
the second register $m$, also of n qbits, contains the memory $|m\rangle$; finally there is a control
register $c$ with $b$ qbits all initialized in the state $|0\rangle$. The full initial quantum state is thus:
\begin{equation}
|\psi_0\rangle = {1\over \sqrt{p}} \sum_{k=1}^p |i; p^k;
0_1,\dots , 0_b\rangle
\label{ab}
\end{equation}
where $|i\rangle = |i_1,\dots ,i_n\rangle$ denotes the input qbits, the second register, $m$, contains the memory (\ref{a})
and all $b$ control qbits are in state $|0\rangle$. Applying the Hadamard gate to the first control qbit one obtains
\begin{eqnarray}
|\psi _1\rangle &&= {1\over \sqrt{2p}} \ \sum_{k=1}^p |i; 
p^k; 0_1,\dots ,0_b\rangle
\nonumber \\ 
&&+{1\over \sqrt{2p}} \ \sum_{k=1}^p |i; 
p^k; 1_1,\dots ,0_b\rangle \ .
\label{ad}
\end{eqnarray}
Let us now apply to this state the following combination of quantum gates:
\begin{equation}
|\psi _2\rangle = \prod_{j=1}^n \ NOT_{m_j} 
\ XOR_{i_j m_j} |\psi _1\rangle \ ,
\label{ae}
\end{equation}
As a result of the above operation the memory register qbits are in state $|1\rangle$ if $i_j$ and $p^k_j$ are identical
and $|0\rangle$ otherwise:
\begin{eqnarray}
|\psi _2\rangle &&= {1\over \sqrt{2p}} \ \sum_{k=1}^p |i; 
d^k; 0_1,\dots ,0_b\rangle 
\nonumber \\
&&+{1\over \sqrt{2p}} \ \sum_{k=1}^p |i; 
d^k; 1_1,\dots ,0_b\rangle \ ,
\label{af}
\end{eqnarray}
where $d^k_j = 1$ if and only if  $i_j=p^k_j$ and $d^k_j=0$ otherwise.

Consider now the following Hamiltonian:
\begin{eqnarray}
{\cal H} &&= \left( d_H \right)_m \otimes \left( \sigma^z \right)_{c_1} \ ,
\nonumber \\
\left( d_H \right)_m && = \sum_{j=1}^n 
\left( {\sigma^z + 1\over 2} \right) _{m_j}\ ,
\label{ag}
\end{eqnarray}
where $\sigma^z$ is the third Pauli matrix.
${\cal H}$ measures the number of 0's in register $m$, with a plus sign if $c_1$ is in state $|0\rangle$ and a minus sign if $c_1$ is in state $|1\rangle$. Given how we have prepared the state $|\psi _2\rangle$, this is nothing else than the number of qbits which are different in the input and memory registers $i$ and $m$. This quantity is called the {\it Hamming distance} and represents the
(squared) Euclidean distance between two binary patterns. 

Every term in the superposition (\ref{af}) is an eigenstate of ${\cal H}$ with a different eigenvalue. Applying thus the unitary operator ${\rm exp} (i \pi {\cal H}/2n)$ to $|\psi _2\rangle$ one obtains
\begin{eqnarray}
|\psi _3\rangle &&= {\rm e}^{i{\pi \over 2n}{\cal H}} \ |\psi_2\rangle \ ,
\label{ah} \\
|\psi_3\rangle &&= {1\over \sqrt{2p}} \sum_{k=1}^p {\rm e}^{i{\pi\over 2n}
d_H\left( i, p^k\right)}
|i; d^k; 0_1,\dots ,0_b\rangle 
\nonumber \\
&&+ {1\over \sqrt{2p}} \sum_{k=1}^p {\rm e}^{-i{\pi\over 2n}
d_H\left( i, p^k\right)}
|i; d^k; 1_1,\dots ,0_b\rangle \ ,
\nonumber
\end{eqnarray}
where $d_H\left( i, p^k \right)$ denotes the Hamming distance bewteen the input $i$ and the stored pattern $p^k$. 

In the final step we restore the memory gate to the state $|m\rangle$ by applying the inverse transformation to eq. (\ref{ae}) and we apply the Hadamard gate to the control qbit $c_1$, thereby obtaining
\begin{eqnarray}
|\psi _4\rangle &&= H_{c_1} \prod_{j=n}^1 XOR_{i_j m_j}  
\ NOT_{m_j} \ |\psi_3\rangle \ ,
\label{ai} \\
|\psi_4\rangle &&= {1\over \sqrt{p}} \sum_{k=1}^p {\rm cos}\  {\pi \over
2n} d_H\left( i, p^k\right) 
|i; p^k; 0_1,\dots ,0_b\rangle 
\nonumber \\
&&+ {1\over \sqrt{p}} \sum_{k=1}^p {\rm sin}\  {\pi \over 2n}
d_H\left( i, p^k\right) 
|i; p^k; 1_1,\dots ,0_b\rangle .
\nonumber
\end{eqnarray}

The idea is now to repeat the above operations sequentially for all $b$ control qbits $c_1$ to $c_b$. This gives 
\begin{eqnarray}
|\psi_{\rm fin}\rangle &&= {1\over \sqrt{p}} \sum_{k=1}^p \sum_{l=0}^b
\ {\rm cos}^{b-l} \left( {\pi\over 2n} d_H\left( i, p^k \right)\right) \times 
\nonumber \\
&&{\rm sin}^l \left( {\pi\over 2n} d_H\left( i, p^k \right)\right) 
\ \sum_{\left\{ J^l \right\}} |i; p^k; J^l\rangle ,
\label{al}
\end{eqnarray}
where $\left\{ J^l \right\}$ denotes the set of all binary numbers of $b$ bits with exactly $l$ bits 1 and $(b-l)$ bits 0.

Note that one could also dispense with a register for the input but, rather, code also the input directly into a unitary operator. 
Indeed, the auxiliary quantum register for the input is needed only by the operator (\ref{ae}) leading from (\ref{ad}) to
(\ref{af}). The same result (apart from an irrelevant overall sign) can be obtained by applying
\begin{eqnarray}
I &&= \prod_{j=1}^n U_j \ ,
\nonumber \\
U_j &&= {\rm sin}\left( {\pi \over 2} i_j \right) 1 + i\ {\rm cos}
\left( {\pi\over 2} i_j \right) \sigma_2 \ ,
\label{newg}
\end{eqnarray}
directly on the memory state $|m\rangle$.  The rest of the algorithm is the same, apart the reversing of the operator (\ref{ae}) which needs now the operator $I^{-1}$. 

The end effect of the information retrieval algorithm represents thus a rotation of the memory quantum state in the enlarged Hilbert space obtained by adding $b$ control qbits. The overall effect of this rotation is an amplitude concentration on memory states similar to the input, if there is a large number of $|0\rangle$ control qbits in the output state and an amplitude concentration on states different from the input, if there is a large number of $|1\rangle$ control qbits in the output state. As a consequence, the most interesting state for information retrieval purposes is the projection of $|\psi_{\rm fin}\rangle$ onto the subspace with all control qbits in state $|0\rangle$.

There are two ways of obtaining this projection. The first, and easiest one, is to simply repeat the above algorithm and measure the control register several times, until exactly the desired state for the control register is obtained. If the number of such repetitions exceeds a preset threshold $T$ the input is classified as "non-recognized" and the algorithm is stopped. Otherwise, once $|c_1, \dots , c_b\rangle = |0_1, \dots, 0_b\rangle$ is obtained, one proceeds to a measurement of the memory register $m$, which yields
the output pattern of the memory. 

The second method is to first apply $T$ steps of the amplitude amplification algorithm \cite{amam} rotating $|\psi_{\rm fin}\rangle$ towards its projection onto the "good" subspace formed by the states with all control qbits in state $|0\rangle$. To this end it is best to use the version of the retrieving algorithm that does not need an auxiliary register for the input. Let us define as $R(i)$ the input-dependent operator which rotates the memory state in the Hilbert space enlarged by the $b$ control qbits towards the
final state $|\psi _{\rm fin}\rangle$ in eq. (\ref{al}) (where we now omit the auxiliary register for the input): 
\begin{equation}
|\psi_{\rm fin}\rangle = R(i) \ |m;0_1,\dots,0_b\rangle \ .
\label{newh}
\end{equation}
By adding also the two utility qbits needed for the storing algorithm one can then obtain $|\psi_{\rm fin}\rangle$ as a unitary transformation of the initial state with all qbits in state $|0\rangle$:
\begin{equation}
|\psi_{\rm fin};00\rangle = R(i) M \ |0,\dots,0;0_1,\dots,0_b;00\rangle \ .
\label{newi}
\end{equation}
The amplitude amplification rotation of $|\psi_{\rm fin};00\rangle$ towards its "good" subspace in which all $b$ control qbits are in state $|0\rangle$ is then obtained \cite{amam} by repeated application of the operator 
\begin{equation}
Q = - R(i)MS_0 M^{-1}R^{-1}(i)S 
\label{newl}
\end{equation}
on the state $|\psi_{\rm fin};00\rangle$. Here $S$ conditionally changes the sign of the amplitude of the "good" states with the $b$ control qbits in state $|0\rangle$, while $S_0$ changes the sign of the amplitude if and only if the state is the zero state $|0,\dots,0;0_1,\dots, 0_b;00\rangle$. As before, if a measurement of the control register after the $T$ iterations of the amplitude amplification rotation yields $|0_1, \dots, 0_b\rangle$ one proceeds to a measurement of the memory register, otherwise the input is classified as "non-recognized". 

The expected number of repetitions needed to measure the desired control register state is $1/P_b^{\rm rec}$, with
\begin{equation}
P_b^{\rm rec} = {1\over p} \ \sum_{k=1}^p \ {\rm cos}^{2b} \left(
{\pi \over 2n} d_H \left( i; p^k\right) \right) 
\label{am}
\end{equation}
the probability of measuring $|c_1,\dots ,c_n\rangle = |0_1, \dots ,0_n\rangle$. The threshold $T$ governs thus the {\it recognition efficiency} of the memory. Note, however, that amplitude amplification provides a quadratic boost \cite{amam} to the recognition efficiency since only $1/\sqrt{P_b^{\rm rec}}$ steps are typically required to rotate $|\psi_{\rm fin}\rangle$ onto the desired subspace. Accordingly, the threshold $T$ can be lowered to $\sqrt{T}$ with respect to the method of projection by measurement. The crucial point is that, due to the quantum nature of the retrieval mechanism, this recognition probability depends on the distribution of {\it all} stored patterns. A lower bound on the recognition probability can thus be established as follows. Of all the stored patterns, all but one have Hamming distance from the input smaller or equal than $(n-1)$. There is only pattern that can have a larger Hamming distance equal to $n$. So we shall use the upper bound $(n-1)$ for the Hamming distance of all patterns but one, for which we shall use the upper bound $n$, and this one does not contribute to the recognition probability since the cosine function vanishes. Given that cosine is a decreasing function in the interval $[0, \pi/2]$, we get the lower bound
\begin{equation}
P_b^{\rm rec} \ge P_b^{\rm min} = {p-1\over p} \ {\rm cos}^{2b} \left( {\pi (n-1) \over 2n} \right) \ .
\label{lowbound}
\end{equation}
For $n \gg 1$ we can now estimate this lower bound as 
\begin{equation}
P_b^{\rm min} \simeq {p-1\over p} \  \left( {\pi \over 2n} \right)^{2b} \ .
\label{estimate}
\end{equation}
This shows that, independent of the number $p$ of patterns, the threshold $T$ for recognition can be set as a polynomial function of the number $n$ of qubits. Note that this is entirely due to the factor $(p-1)$ in the numerator of (\ref{estimate}), which, in turn, depends on the quantum nature of the memory. In other words, the probabilistic character of the retrieval process does not limit at all the number of possible stored patterns, the typical number of repetitions required would be polynomial even for an exponential number or patterns. The efficient implementability of the quantum memory is limited only by the number of elementary quantum gates in $M$, which is linear in $p$. 

In general, the probability of recognition is determined by comparing (even) powers of cosines and sines of the distances to the stored patterns. It is thus clear that the worst case for recognition is the situation in which there is an isolated pattern,
with the remaining patterns forming a tight cluster spanning all the largest distances to the first one. As a consequence,
the threshold needed to recognize all patterns diminishes when the number of stored patterns becomes very large, since, in this case, the distribution of patterns becomes necessarily more homogeneous. Indeed, for the maximal number of stored patterns
$p=2^n$ one has $P_b^{\rm rec}= 1/2^b$ and the recognition efficiency becomes also maximal, as it should be. 

Once the input pattern $i$ is recognized, the measurement of the memory register yields the stored pattern $p^k$ with probability
\begin{eqnarray}
P_b\left( p^k\right) &&= {1\over Z} \ {\rm cos}^{2b} \left( {\pi \over 2n}
d_H\left( i, p^k\right) \right) \ ,
\label{an} \\
Z &&= pP_b^{\rm rec} = \sum_{k=1}^p {\rm cos}^{2b} \left( {\pi \over 2n}
d_H\left( i, p^k\right) \right) \ .
\label{ao}
\end{eqnarray}
Clearly, this probability is peaked around those patterns which have the smallest Hamming distance to the input. The highest probability of retrieval is thus realized for that pattern which is most similar to the input. This is always true, independently of the number of stored patterns. In particular, contrary to classical associative memories, there are {\it no spurious memories}: the probability of obtaining as output a non-stored pattern is always zero. This is another manifestation of the fact that there are no restrictions on the loading factor $p/n$ due to the information retrieval algorithm.

In addition to the threshold $T$, there is a second tunable parameter, namely the number $b$ of control qbits. This new parameter $b$ controls the {\it identification efficiency} of the quantum memory since, increasing $b$, the probability distribution $P_b\left( p^k\right)$ becomes more and more peaked on the low $d_H\left( i, p^k \right) $ states, until
\begin{equation}
\lim_{b\to \infty} P_b\left( p^k \right) = \delta_{k k_{\rm min}}\ ,
\label{ap}
\end{equation}
where $k_{\rm min}$ is the index of the pattern (assumed unique for convenience) with the smallest Hamming distance to the input.

While the recognition efficiency depends on comparing powers of cosines and sines of the same distances in the distribution, the
identification efficiency depends on comparing the (even) powers of cosines of the different distances in the distribution.
Specifically, it is best when one of the distances is zero, while all others are as large as possible, such that the probability of retrieval is completely peaked on one pattern. As a consequence, the identification efficiency is best when the recognition efficiency is worst and viceversa.  

The role of the parameter $b$ becomes familiar upon a closer examination of eq.( \ref{an}). Indeed, the quantum distribution described by this equation is equivalent to a canonical Boltzmann distribution with (dimensionless) temperature
$t = 1/b$ and (dimensionless) energy levels 
\begin{equation}
E^k = -2 \ {\rm log} \ {\rm cos} \left( {\pi \over 2n} d_H\left( i, p^k
\right) \right) \ ,
\label{aq}
\end{equation}
with $Z$ playing the role of the partition function. 

The appearance of an effective thermal distribution suggests studying the average behaviour of quantum associative memories via the corresponding thermodynamic potentials. Before this can be done, however, one must deal with the different distributions of stored patterns characterizing each individual memory. The standard way to do this in similar classical problems is to average over the random distribution of patterns. Typically, one considers quenched averages in which extensive quantities, like the free energy  are averaged over the disorder: this is the famed replica trick used to analyze spin glasses \cite{parisi}. In the present case, however, the disorder cannot lead to spin-glass-like phases since there are no spurious memories: by construction, probabilistic quantum memories can output only one of the stored patterns. The only question is how accurate is the retrieval of the most similar pattern to the input as a function of the fictitious temperature $t=1/b$. To address this question we will "quench" only one aspect of the random pattern distribution, namely the minimal Hamming distance $d$ between the input and the stored patterns. The rest of the random pattern distribution will be considered as annealed. In doing so, one obtains an average description of the average memory as a function of the fictitious temperature $t=1/b$ and the minimal Hamming distance $d$.

To do so we first normalize the pattern representation by adding (modulo 2) to all patterns, input included, the input pattern $i$. This clearly preserves all Hamming distances and has the effect of choosing the input as the state with all qbits in state $|0\rangle$. The Hamming distance $d_H\left( i, p^k\right)$ becomes thus simply the number of qbits in pattern $p^k$ with value $|1\rangle$. The averaged partition function takes then a particularly simple form:
\begin{equation}
Z_{\rm av} = {p\over N_{\lambda}} 
\ \sum_{\{\lambda \}} \ \sum_{j=d}^{n} \ \lambda_j
\ {\rm cos}^{2b}\left( {\pi \over 2} {j\over n}\right) \ ,
\label{ar}
\end{equation}
where $\lambda_j$ describes a probability distribution, $\sum_{j=d}^n \lambda_j = 1$, with the following properties. Let the number of patterns scale as the ${\rm x}^{\rm th}$ power of the number of qubits, $p=\alpha_x n^x$ for $n\gg 1$. Then
\begin{eqnarray}
\lambda_j &&= 0 \ , \qquad \qquad j > n-x \ ,
\nonumber \\
\lambda_j &&\le {1\over \alpha_x x!}\ , \ \qquad j=n-x \ ,
\label{condit}
\end{eqnarray}
with all other $\lambda_j$ for $j < n-x$ unconstrained.  $\{ \lambda \}$ is the set of such distributions and $N_{\lambda}$ the corresponding normalization factor. Essentially the probability distribution becomes unconstrained in the limit of large $n$. 

We now introduce the free energy $F(b,d)$ by the usual definition 
\begin{equation}
Z_{\rm av} = p \ {\rm e}^{-bF(b,d)} = Z_{\rm av}(b=0) \ {\rm e}^{-bF(b,d)} \ ,
\label{as}
\end{equation}
where we have chosen a normalization such that ${\rm exp}(-bF)$ describes the deviation of the partition function from its value for $b=0$ (high effective temperature). Since $Z/p$, and consequently also $Z_{\rm av}/p$  posses a finite, non-vanishing large-$n$ limit, this normalization ensures that $F(b,d)$ is intensive, exactly like the energy levels (\ref{aq}), and scales as a constant for large $n$. This is the only difference with respect to the familiar situation in statistical mechanics.

The free energy describes the equilibrium of the system at effective temperature $t=1/b$ and has the usual expression in terms of the internal energy $U$ and the entropy $S$:
\begin{eqnarray}
F(t,d) &&= U(t,d) - tS(t,d) \ ,
\nonumber \\
U(t,d) && = \langle E \rangle _t \ ,\quad 
S(t,d) = {-\partial F(t,d) \over \partial t} \ .
\label{at}
\end{eqnarray}
Note that, with the normalization we have chosen in (\ref{as}), the entropy $S$ is always a negative quantity describing the deviation from its maximal value $S_{\rm max} = 0$ at $t=\infty$.

By inverting eq.(\ref{aq}) with $F$ substituting $E$ one can also define an effective (relative) input/output Hamming distance ${\cal D}$ at temperature $t$:
\begin{equation}
{\cal D}(t,d) = {2\over \pi} \ {\rm arccos} \ {\rm e}^{-F(t,d)\over 2} \ .
\label{au}
\end{equation}
This corresponds exactly to representing the recognition probability of the average memory as
\begin{equation}
\left( P_b^{\rm rec} \right) _{\rm av} = 
{\rm cos}^{2b} \left( {\pi \over 2} {\cal D}(b,d) \right) \ ,
\label{av}
\end{equation}
which can also be taken as the primary definition of the effective Hamming distance. 

The function ${\cal D}(b,d)$ provides a complete description of the behaviour of the average probabilistic quantum associative memory  with a minimal distance Hamming distance $d$. 
This can be used to tune its performance. Indeed, suppose that one wants the memory to recognize and identify
inputs with up to $\epsilon n$ corrupted inputs with an efficiency of $\nu$ $(0\le \nu \le 1)$. Then one must choose a number $b$ of control qbits sufficiently large that 
$\left( {\cal D}(b,\epsilon n) - \epsilon \right) \le \left( 1-\nu \right)$ and a threshold $T$ of repetitions satisfying $T \ge 1/{\rm cos}^{2b} \left({\pi \over 2} {\cal D}(b,\epsilon n) \right) $, as illustrated in Fig. 2 below.

A first hint about the general behaviour of the effective distance function ${\cal D}(b,d)$ can be obtained by examining closer the energy eigenvalues (\ref{aq}). For small Hamming distance to the input these reduce to
\begin{equation}
E^k \simeq {\pi^2\over 4} \left( {d_H\left( i, p^k\right) \over n}
\right) ^2\ ,\qquad {d_H\left( i, p^k \right) \over n} \ll 1\ .
\label{aw}
\end{equation}
Choosing again the normalization in which $|i\rangle = |0 \dots 0\rangle$ and introducing a ``spin" $s_i^k$ with value $s_i^k = -1/2$ if qbit $i$ in pattern $p^k$ has value $|0\rangle$ and $s_i^k=+1/2$ if qbit $i$ in pattern $p^k$ has value $|1\rangle$, one can express the energy levels for $d_H/n \ll 1$ as
\begin{equation}
E^k = {\pi ^2\over 16} + {\pi ^2\over 4n^2}
\sum_{i,j} s_i^k s_j^k + {\pi ^2 \over 4n} \sum_i s_i^k \ .
\label{ay}
\end{equation}
Apart from a constant, this is the Hamiltonian of an infinite-range antiferromagnetic Ising model in presence of a magnetic field. The
antiferromagnetic term favours configurations $k$ with half the spins up and half down, so that $s^k_{\rm tot} =\sum_i s^k_i = 0$, giving $E^k =\pi^2/16$.The magnetic field, however, tends to align the spins so that $s^k_{\rm tot} = -n/2$, giving $E^k = 0$. Since this is lower than $\pi^2/16$, the ground state configuration is ferromagnetic, with all qbits having value $|0\rangle$. At very low temperature (high $b$), where the energy term dominates the free energy, one expects thus an ordered phase of the
quantum associative memory with ${\cal D}(t,d) = d/n$. This corresponds to a perfect identification of the presented input. As the temperature is raised ($b$ decreased) however, the thermal energy embodied by the entropy term in the
free energy begins to counteract the magnetic field. At very high temperatures (low $b$) the entropy approaches its maximal value $S(t=\infty) = 0$ (with the normalization chosen here). If this value is approached faster than $1/t$, the free energy will again be dominated by the internal energy . In this case, however, this is not any more determined by the ground state but rather equally
distributed on all possible states, giving
\begin{eqnarray}
F(t=\infty) &&= U(t=\infty) = {-1\over 1-{d\over n}} 
\int_{d\over n}^1 \ dx
\ 2\ {\rm log} \ {\rm cos}\left( {\pi \over 2} x\right) 
\nonumber \\
&&= \left( 1+{d\over n}
\right) 2 \ {\rm log }2 + O\left( \left( {d\over n}\right) ^2\right) \ ,
\label{az}
\end{eqnarray}
and leading to an effective distance
\begin{equation}
{\cal D}(t=\infty, d) = {2\over 3} - {2\ {\rm log }2 \over \pi \sqrt{3}} 
\ {d\over n} + O\left( \left( {d\over n}\right) ^2\right) \ .
\label{azz}
\end{equation}
This value corresponds to a disordered phase with no correlation between input and output of the memory. 

A numerical study of the thermodynamic potentials in (\ref{at}) and (\ref{au}) indeed confirms a phase transition from the ordered to the disordered phase as the effective temperature is raised. In Fig. 2 we show the effective distance ${\cal D}$ and the entropy $S$ 
for 1 Mb ($n=8 \times 10^6$) patterns and $d/n= 1\%$ as a function of the inverse temperature $b$ (the entropy is rescaled
to the interval [0,1] for ease of presentation). At high temperature there is indeed a disordered phase with $S=S_{\rm max} =0$ and ${\cal D} = 2/3$. At low temperatures, instead, one is in the ordered phase with $S=S_{\rm min}$ and
${\cal D}=d/n=0.01$. The effective Hamming distance plays thus the role of the order parameter for this quantum phase transition. 

\begin{figure}
\includegraphics[width=14cm]{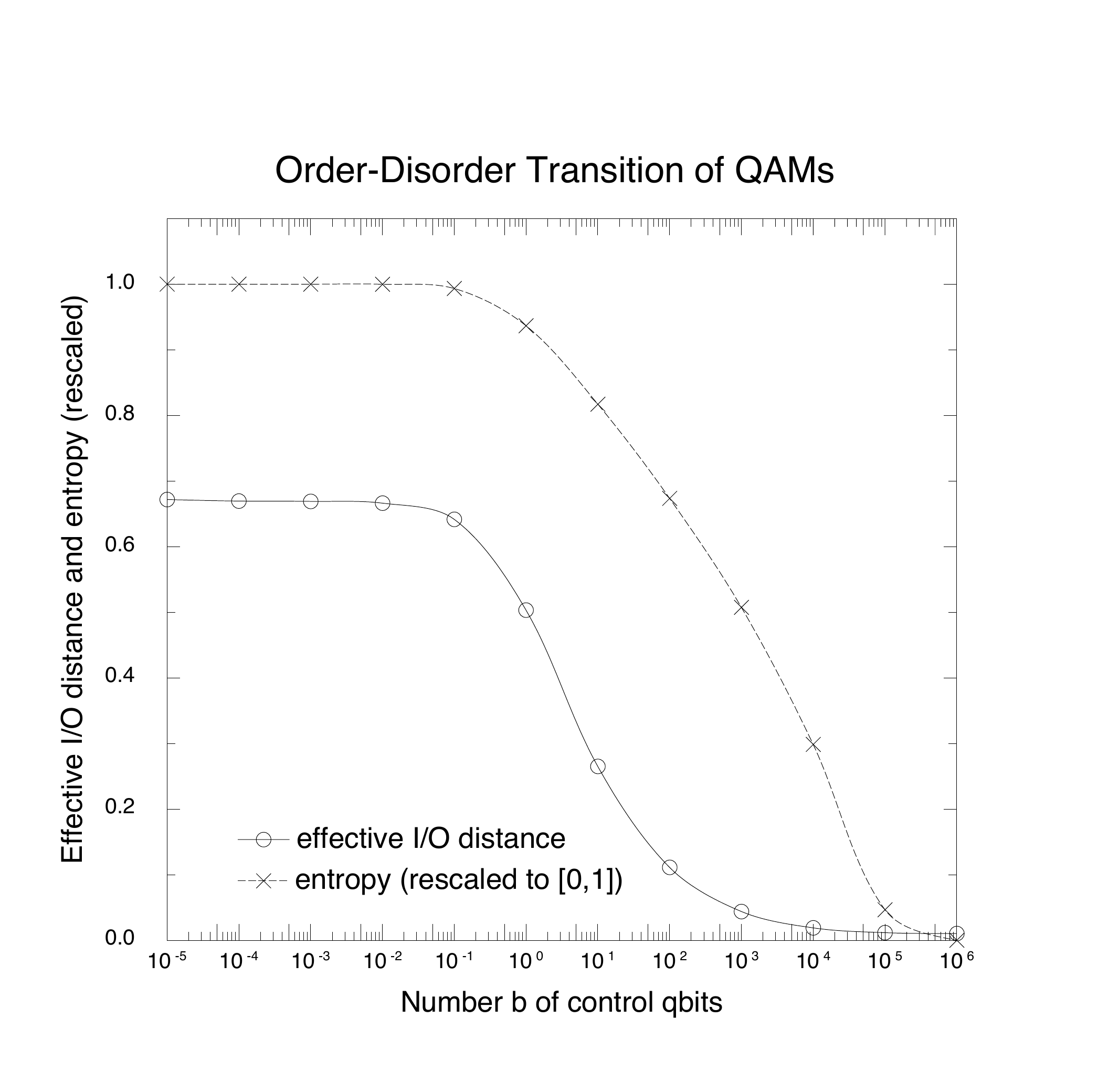}
\caption{Effective input/output distance and entropy (rescaled to [0,1]) 
for 1Mb patterns and $d/n = 1\%$.}
\end{figure}

The phase transition occurs around $b_{\rm cr} \simeq 10^{-1}$. The physical regime of the quantum associative memory ($b$ = positive integer) lies thus just above this transition. For a good accuracy of pattern recognition one should choose a fictitious temperature low enough to be well into the ordered phase.  As is clear from Fig. 2, this can be achieved already with a number of control qubits $b=O(10^4)$. 

Having described at length the information retrieval mechanism for complete, but possibly corrupted patterns, it is easy to incorporate also incomplete ones. To this end assume that only $q<n$ qbits of the input are known and let us denote these by the indices $\{k1, \dots, kq\}$. After assigning the remaining qbits randomly, there are two possibilities. One can just treat the resulting complete input as a noisy one and proceed as above or, better, one can limit the operator $\left( d_H \right)_m$ in the 
Hamiltonian (\ref{ag}) to
\begin{equation}
\left( d_H \right)_m = \sum_{i=1}^q \ \left( {\sigma^z+1\over 2} 
\right)_{m_{ki}} \ ,
\label{p}
\end{equation}
so that the Hamming distances to the stored patterns are computed on the basis of the known qbits only. After this, the pattern recall process continues exactly as described above. This second possibility has the advantage that it does not introduce random noise in the similarity measure but it has the disadvantage that the operations of the memory have to be adjusted to the inputs.

Finally, it is fair to mention that the model of probabilistic quantum associative memory presented here has been criticised \cite{zalka} on three accounts:
\begin{itemize}
\item{} It has been claimed that the same result could have been obtained by storing only one of the $p$ patterns in $n$ classical bits and always using this single pattern as the same output independently of the input, provided the input has a Hamming distance to the unique stored pattern lower than a given threshold, otherwise the input would not be recognized.
\item{} It has been claimed that the Holevo theorem bounds the number of patterns that can be stored in a quantum associative memory.
\item{} It has been pointed out that the complexity of memory preparation prevents the efficient storing of patterns. 
\end{itemize}
This criticism is wrong on the first two accounts and partially justified on the third \cite{reply}. It is true that both the quantum memory and the proposed equivalent classical prescription are based on probabilistic recognition and identification processes. In the proposed  classical alternative, however the probabilities for both recognition and identification depend on {\it one unique, fixed and random pattern} whereas in the quantum memory, exactly due to its quantum character, these probabilities depend on {\it the distribution of all stored patterns}. These probabilities are such that an input different from most stored patterns is more difficult to recognize than an input similar to many stored memories and that the identification probability distribution can be peaked with any prescribed accuracy on the stored pattern most similar to the input. In the proposed classical alternative, given that only one single pattern can be stored on the $n$ classical bits, the recognition or lack thereof depend on the distance to a randomly chosen pattern and the identification probability is a delta function peaked on this fixed random pattern. In other words there is no correlation whatsoever between input and output apart from the fact that they have Hamming distance below a certain threshold, a prescription that can hardly qualify as an associative memory: it would indeed be a boring world the one in which every stimulus would produce exactly the same response, if any response at all. Also, the Holevo theorem \cite{holevo} does not impose any limitation on this type of probabilistic quantum memories. The Holevo theorem applies to the situation in which Alice codes information about a classical random variable in a quantum state and Bob tries to retrieve the value of this random variable by measurements on the received quantum state. In the present case Alice gives to Bob also corrupted or incomplete {\it classical information} about the random variable (the input) and Bob can use also a {\it unitary transformation that encodes both the memories and the input} (operator $R(i)$ in (\ref{newh})) in addition to measurements, a completely different situation. Contrary to what the authors of \cite{zalka} affirm, a memory that "knows the patterns it is supposed to retrieve" not only makes sense but it is actually the very definition of an associative memory: if the memory would not "know" the data it has to retrieve it would just be a random access database, exactly the architecture that one wants to improve by content association, the mechanism whose goal is to recognize and correct corrupted or incomplete inputs. The dynamics of the classical Hopfield model "knows" the patterns it is supposed to retrieve: they are encoded in the neuronal weights. So does any human brain. Finally, the third critique is partially correct. The complexity of the memory operator $M$ is $O(pn)$ and thus the original claim \cite{tru1} of an exponential capacity gain by quantum associative memories is excessive. This, however, does not invalidate the main claim, a large gain in capacity is made possible by quantum mechanics, albeit only a polynomial one. This correction has been incorporated in the present review.

\subsection{Efficiency, complexity and memory tuning}
In this last section we would like to address the efficient implementation of probabilistic quantum memories in the quantum circuit model \cite{review} and their accuracy tuning. 

We have stressed several times that all unitary operators involved in the memory preparation can be realized as a sequence of one- and two-qubit operators. It remains to prove that this is true also for pattern retrieval and that all these operators 
can be implemented in terms of a small set of universal gates. To this end we would like to point out that, in addition to the standard NOT, H (Hadamard), XOR, 2XOR (Toffoli) and nXOR gates \cite{review} we have introduced only the two-qbit gates $CS^i$ in eq. (\ref{addb}) and the unitary operator ${\rm exp}\left( i\pi {\cal H}/2n \right)$. The latter can, however also be realized by simple gates involving only one or two qbits. To this end we introduce the single-qbit gate
\begin{equation}
U = \left( \begin{matrix} {\rm e}^{i{\pi\over 2n}}&0\\
0&1\\ \end{matrix} \right) \ ,
\label{q}
\end{equation}
and the two-qbit controlled gate
\begin{equation}
CU^{-2} = |0\rangle \langle 0| \otimes 1 + |1\rangle \langle 1|
\otimes U^{-2} \ .
\label{r}
\end{equation}
It is then easy to check that ${\rm exp}\left( i\pi {\cal H}/2n \right)$ 
in eq. (\ref{af}) can be realized as follows:
\begin{equation}
{\rm e}^{i{\pi \over 2n} {\cal H}} \ |\psi _2\rangle = 
\prod_{i=1}^n \left( CU^{-2} \right) _{c m_i} \ \prod_{j=1}^n U_{m_j}
\ |\psi_2\rangle \ ,
\label{t}
\end{equation}
where $c$ is the control qbit for which one is currently repeating the algorithm. Essentially, this means that one implements first ${\rm exp}\left( i\pi  d_H /2n \right)$ and then one corrects by implementing ${\rm exp}\left( -i\pi  d_H /n \right)$ on that part of
the quantum state for which the control qbit $|c\rangle$ is in state $|1\rangle$. 

Using this representation for the Hamming distance operator one can count the total number of simple gates that one must apply in order to implement one step of the information retrieval algorithm. This is given by $(6n+2)$ using the auxiliary register for the input and by $(4n+2)$ otherwise. This retrieval step has then to be repeated for each of the $b$ control qbits. Therefore,
implementing the projection by repeated measurements, the overall complexity $C$ of information retrieval is bounded by 
\begin{equation}
C \le T b (6n+2) C_{\rm M}\ ,
\label{neweqe}
\end{equation}
where $C_{\rm M}$ is the complexity of the memory preparation, given by the operator $M$ or a probabilistic cloning machine. In particular, it is given by
\begin{equation}
C =T b (6n+2) \left( p(2n+3)+1 \right) \ ,
\label{neweqeadd}
\end{equation}
for the simplest version of the algorithm, using memory preparation by $M$ and an auxiliary input register. 

The computation of the overall complexity is easier for the information retrieval algorithm which uses the amplitude amplification technique. In this case the initial memory is prepared only once by a product of the operators $M$, with complexity $p(2n+3)+1$ and $R(i)$, with complexity $b(4n+2)$. Then one applies
$T$ times the operator $Q$, with complexity $p(4n+6) + b(8n+4) + 2 + C_S + C_{S_0}$, where $C_S$ and $C_{S_0}$ are the polynomial  complexities of the oracles implementing $S$ and $S_0$. This gives
\begin{eqnarray}
C &&= T\left[ p(4n+6) + b(8n+4) + 2 +C_S + C_{S_0}\right] +
\nonumber \\
&&+ p(2n+3) + b(4n+2)+1\ .
\label{newm}
\end{eqnarray}
As expected, the memory complexity (be it (\ref{neweqeadd}) or (\ref{newm})) depends on both $T$ and $b$, the parameters governing the recognition and identification efficiencies. The major limitation comes from the factor $p$ representing the total number of stored patterns. Note however that, contrary to classical associative memories, one can efficiently store and retrieve any polynomial number of patterns due to the absence of spurious memories and crosstalk. 

Let us finally show how one can tune the accuracy of the quantum memory. Suppose one would like to recognize on average inputs with up to 1\% of corrupted or missing bits and identify them with high accuracy. The effective
i/o Hamming distance ${\cal D}$ shown in Fig. 2 can then be used to determine the values of the required parameters $T$ and $b$ needed to reach this accuracy for the average memory.  For $b=10^4$ e.g., one has ${\cal D} = 0.018$, which  gives the average i/o distance (in percent of total qbits) if the minimum possible i/o distance is 0.01. For this value of $b$ the recognition probability
is $3.4\ 10^{-4}$. With the measurement repetition technique one should thus set the threshold $T \simeq 3000 $. Using amplitude amplification, however, one needs only around $T=54$ repetitions. Note that the values of $b$ and $T$ obtained by tuning the memory with the effective i/o Hamming distance become $n$-independent for large values of $n$. This is because they are intensive variables unaffected by this "thermodynamic limit". For any fixed $p$ polynomial in $n$, the information retrieval can then be implemented efficiently and the overall complexity is determined by the accuracy requirements via the n-independent parameters $T$ and $b$.

\section{Conclusion}
We would like to conclude this review by highlighting the fundamental reason why a probabilistic quantum associative memory works better than its classical counterpart and pointing out about some very intuitive features of the information retrieval process. 

In classical associative memories, the information about the patterns to recall is typically stored in an energy function. When retrieving information, the input configuration evolves to the corresponding output, driven by the dynamics associated with the
memory function. The capacity shortage is due to a phase transition in the statistical ensemble governed by the memory energy function. Spurious memories, i.e. spurious metastable minima not associated with any of the original patterns become important for loading factors $p/n$ above a critical value and wash out completely the memory, a phenomenon that goes by the name of crosstalk. So, in the low $p/n$ phase the memory works perfectly in the sense that it outputs always the stored
pattern which is most similar to the input. For $p/n$ above the critical value, instead, there is an abrupt transition to total amnesia caused by spurious memories. 

Probabilistic quantum associative memories work better than classical ones since they are {\it free from spurious memories}. The easiest way to see this is in the formulation
\begin{equation}
|m\rangle = M \ |0\rangle \ .
\label{newn}
\end{equation}
All the information about the stored patterns is encoded in the unitary operator $M$. This generates a quantum state in which all components that do not correspond to stored patterns have exactly vanishing amplitudes.

An analogy with the classical Hopfield model \cite{neuralnetworks} can be established as follows. Instead of generating the memory state $|m\rangle $ from the initial zero state $|0\rangle $, one can start from a uniform superposition of the computational
basis. This is achieved by the operator $MW$ defined by 
\begin{eqnarray}
|m\rangle &&= M W  \ {1\over \sqrt{2^n}} \sum_{j=0}^{2^n-1} |j\rangle \ ,
\nonumber \\
W &&\equiv \prod_{j=1}^n H_j \ .
\label{newo}
\end{eqnarray}
Now, this same result can also be obtained by Grover's algorithm, or better by its generalization with zero failure rate \cite{long}. Here the state $|m\rangle$ is obtained by applying to the uniform superposition of the
computational basis q times the search operator $X$ defined in
\begin{eqnarray}
|m\rangle &&= X^q \ {1\over \sqrt{2^n}} \sum_{j=0}^{2^n-1} |j\rangle \ ,
\nonumber \\
X &&\equiv -W J_0 W J \ ,
\label{newp}
\end{eqnarray}
where $J$ rotates the amplitudes of the states corresponding to the patterns to be stored by a phase $\phi$ which is very close to $\pi$ (the original Grover value) for large $n$ and $J_0$ does the same on the zero state. Via the two equations (\ref{newo}) and (\ref{newp}), the memory operator $M$ provides an implicit realization of the phase shift operator $J$. Being a unitary operator, this can always be written as an exponential of an hermitian Hamiltonian ${\cal H}_M$, which is the quantum generalization of a classical energy function. By defining $J\equiv {\rm exp}\left( -i{\cal H}_M\right)$ one obtains an energy operator which is diagonal in the computational basis and such that the patterns to be stored have energy eigenvalues $E=-\phi \simeq -\pi$ while all others have energy eigenvalues $E=0$. This formulation is the exact quantum generalization of the Hopfield model; the important point is that the operator $M$ realizes efficiently a dynamics in which the patterns to be stored are always, for any number $p$ of patterns, 
the exact global minima of a quantum energy landscape, without the appearance of any spurious memories. 

The price to pay is the probabilistic nature of the information retrieval mechanism. As always in quantum mechanics, the dynamics determines only the evolution of probability distributions and the probabilistic aspect is brought in by the collapse of this probability distributions upon measurement. Therefore, contrary to the classical Hopfield model in the low $p/n$ phase, one does not always have the absolute guarantee that an input is recognized and identified correctly as the stored pattern most similar to the input, even if this state has the highest probability of being measured. But, after all, this is a familiar feature of the most concrete example of associative memory, our own brain, and should thus not be so disturbing. Indeed, it is not only the probabilistic nature of information retrieval that is reminiscent of the behaviour of the human brain but also the properties of the involved probability distributions. These are such that inputs very similar to a cluster of stored patterns will be much easier to recognize than inputs farther away from all stored memories, although the former situation will lead to a more difficult identification of the most similar memory. Quantum entanglement allows to construct such a probabilistic content association with a very high storage capacity.

\end{document}